\documentstyle[preprint,aps,prb]{revtex}

\begin{document}
\draft
\begin{center}
{\large\bf Eliashberg--type equations for correlated superconductors}\\

\medskip
Karol I. Wysoki\'nski \\

\medskip
Institute of Physics, M. Curie--Sk\l odowska University, \\
ul. Radziszewskiego 10A, Pl--20 031 Lublin, Poland
\end{center}

\begin{abstract}
The derivation of the Eliashberg type equations for
a~superconductor with strong electron correlations and
electron--phonon interaction has been presented. The proper
account of short range Coulomb interactions results in strongly
anisotropic equations. Possible symmetries of the order parameter
include s,p and d-wave. We found the carrier concentration dependence
of the coupling constants corresponding to these symmetries. At low
hole doping the d-wave component is the largest one.
\end{abstract}

\bigskip
\noindent
\section{Introduction}

\bigskip
Recently there appeared in the literature a~number of papers
dealing with the problem of electron--phonon interaction in
strongly correlated superconductors \cite{Kim}$^{-}$\cite{Plakida}. 
It turns
out that interplay between Coulomb and electron--phonon
interactions is very subtle and may result in an unexpected 
behaviour of various quantities \cite{Marsiglio}$^{-}$\cite{Freericks}. 

The main motivation for all these studies come from experiments
on high T$_c$ superconductors which in one or other way have
pointed out the importance of electron--phonon interaction.
There have been found Raman and infrared--active
modes \cite{Chrzan} with strongly temperature dependent linewidth.
Small, but nonzero, isotope shift exponent $\alpha$ ($T_c \propto
M^\alpha$, where $M$ the ion isotope mass) \cite{Muller,Bornem}
is the strong indication of the electron--phonon coupling.
Similar conclusions can be inferred from tunneling and
photoemission \cite{Hinks}, neutron \cite{Mook,Arai},
specific heat \cite{Reeves}, thermal conductivity \cite{Cohn,Klamut}
and other experiments. 

It is the purpose of this paper to derive Eliashberg
type \cite{Eliashberg} equations valid for the strongly correlated
superconductor with electron--phonon interaction. To this end we
assume the validity of Migdal \cite{Migdal} theorem which means
that we assume relatively weak electron--phonon interaction. The
strong correlations described by the Hubbard model will be
treated via slave boson method. The derived equations are
strongly anisotropic and lead to the possibility of various types
of symmetries of order parameter. The relative stability of
various symmetries does depend on the carrier concentration and
other parameters.

In Section II we present the model Hamiltonian for strongly
correlated systems and discuss the application of slave
boson approximation to it in the $U = \infty$ limit, both in mean
field and beyond it. The derivation of the Eliashberg equations
on the real frequency axis is described in considerable detail
in Section III. Section IV contains presentation of results and
discussion. 

\bigskip\bigskip
\noindent
\section{The Model}

\bigskip
We start with the one band Hubbard model with general form of
electron--phonon interaction term
\begin{equation}
H = \sum_{ij\sigma} (t_{ij} - \mu\delta_{ij}) 
\tilde{c}^+_{i\sigma} \tilde{c}_{j\sigma} + 
  \sum_{ijs\sigma}T'^\alpha_{ijs} \, u^\alpha_s
  \tilde{c}^+_{i\sigma} \tilde{c}_{j\sigma} 
 +  U \, \sum_i \tilde{n}_{i\uparrow} \tilde{n}_{i\downarrow}
  +  H_{\rm ph}.
\label{eqno1}
\end{equation}
Here $\tilde{c}^+_{i\sigma}(\tilde{c}_{i\sigma})$ is the
creation (annihilation) operator for a spin $\sigma$ electron
at site $i$ of the lattice, $u^\alpha_i$ is the $\alpha$--th
component of the displacement vector of the ion,
$\tilde{n}_{i\sigma} = \tilde{c}^+_{i\sigma} \tilde{c}_{i\sigma}$,
$\mu$ denotes chemical potential, $t_{ij}$ is the hopping
integral assumed to take on nonzero value $-t$ for $i,j$ being
nearest neighbour sites, $U$ is the Hubbard on--site repulsion
of carriers. $H_{\rm ph}$ denotes the Hamiltonian of the
lattice. The electron--lattice interaction described by the
second term in  (1) has two components
$$
 T'^\alpha_{ijs} = T^\alpha_{ijs} + V^\alpha_{js}\delta_{ij},
$$
the first of which, $T^\alpha_{ijs}$ has been derived from
modulation of the hopping integral \cite{elphon} $t_{ij}$ in the
deformed lattice. It is related to the derivative of $t_{ij}$
taken at equilibrium position of an ion
$$
 T^\alpha_{ijs} = {\partial t_{ij} \over \partial R^\alpha_{ij}}
 \delta_{is} - {\partial t_{ij} \over \partial R^\alpha_{ij}}
 \delta_{js}\,.
$$
The second part of the interaction i.e. the term $V^\alpha_{js}$ is
connected with 
fluctuations of the crystal field. Due to  ionic character of
high temperature superconductors this term is expected to be
more important and has to be taken into account even if we have
assumed the equilibrium value of crystal field $\varepsilon_i =
t_{ii}$ to be zero. The systematic derivation and the discussion
of the electron--ion interaction in context of superconducting
oxides can be found in Ref.\cite{Petru}.

We are interested in the strong correlation limit, characterised
by the large $U$ values ($U \gg  W$, where $W = 8t$ is the
width of electron band in a~two dimensional square lattice). In
this limit and for less than half filled band (i.e. concentration of
carriers $n < 1$) it is very inconvenient for two carriers to
occupy the same site $i$. It means that double occupation of
sites is prohibited and the dynamics of carriers is limited to
empty and singly occupied states. Thus we have to project out
all the doubly occupied states of the system. This procedure is
most conveniently carried out with the help of auxiliary
particles \cite{slave} and in the $U = \infty$ limit. One
rewrites the electron operators in (1) in terms of new
fermion operators $c^+_{i\sigma}(c_{i\sigma})$ and auxiliary
boson operator $b^+_i(b_i)$ as $\tilde{c}^+_{i\sigma}
\tilde{c}_{j\sigma} \rightarrow c^+_{i\sigma}c_{j\sigma} b_i
b^+_j$. The term $U$ may then be dropped out at the expense of
introducing at each site a~constraint
\begin{equation}
 \sum_\sigma c^+_{i\sigma}c_{i\sigma} + b^+_i b_i = 1
\label{eqno2}
\end{equation}
via Lagrange multiplier $\Lambda_i$. The constraint allows for
at most single occupation of each size. 

The Hamiltonian (\ref{eqno1}) then  becomes
\begin{eqnarray}
H & = & \sum_{ij\sigma} t_{ij}c^+_{i\sigma}c_{j\sigma}b_ib^+_j -
 \mu \sum_{i\sigma} c^+_{i\sigma}c_{i\sigma} + 
 \sum_{ijs\sigma\alpha} T'^\alpha_{ijs} u^\alpha_s c^+_{i\sigma}
 c_{j\sigma} b_i b^+_j \nonumber \\
& + & \sum_{is\alpha} V^\alpha_{is} u^\alpha_s
  c^+_{i\sigma}c_{i\sigma} + \sum_i \Lambda_i (\sum_\sigma
  c^+_{i\sigma} c_{i\sigma} + b^+_i b_i - 1) + H_{\rm ph}.
\label{eqno3}
\end{eqnarray}
Note that on--site terms are not modified. It is connected with
the meaning of the boson operators $b^+_i$ as operators creating
empty states at site $i$. The hopping of a~real electron from
occupied state $j$ to an empty state at $i$, described by
term $\tilde{c}^+_{i\sigma}\tilde{c}_{j\sigma}$ in (\ref{eqno1})
consists of motion of a~fermion between these sites
($c^+_{i\sigma}c_{j\sigma}$) and at the same time the empty site
moves from site $i$ to $j$ ($b_ib^+_j$).

It has been shown that mean field description can be obtained by
assuming average values $\langle b_i\rangle  = \langle
b^+_j\rangle = r$ and $\Lambda_i = \Lambda$ at each site. Two
parameters $r$ and $\Lambda$ entering the mean field Hamiltonian
can be choosen so as to make minimal the ground state energy
$E_{GS} = \langle H\rangle$. One gets \cite{slave} 
\begin{eqnarray*} 
&& r^2 = 1 - n   \\
&& -\Lambda = {1 \over N} \sum_{ij\sigma} t_{ij} \langle c^+_{i\sigma}
 c_{j\sigma}\rangle + {1 \over N} \sum_{ijs\alpha\sigma} 
  T'^\alpha_{ijs} \langle u^\alpha_s c^+_{i\sigma}c_{j\sigma}\rangle.
\end{eqnarray*}
Here $n = {1 \over N} \sum_{i\sigma} \langle c^+_{i\sigma} 
c_{\sigma}\rangle$ denotes the concentration of electrons in the
band ($n < 1$).
There are two modifications of the spectrum of electrons encountered
on the mean 
field level. First is the band narrowing described by
$r^2$ an its shift described by $\Lambda$. The
spectrum of noninteracting fermions in the mean field is given
by $(r^2 \epsilon_k - \mu + \Lambda)$ instead of $(\epsilon_k -
\mu)$ of original electrons (at $U = 0$). For the half filled
band $n = 1$ the system is localised ($r = 0$). Here $\epsilon_k$ is
the Fourier transform of $t_{ij}$.

To go beyond mean field approximation for slave bosons it is
convenient to define boson fluctuating fields via: $b^+_i = r +
\delta b^+_i$, $b_i = r + \delta b_i$.

In view of our main goal in next section which is derivation of
the Eliashberg equations, we shall write down the Hamiltonian in
terms of Gorkov--Nambu \cite{Nambu} field operators:
$$
 \psi_i = 
 \left(
 \begin{array}{l}
 c_{i\uparrow} \\
 c^+_{i\downarrow}
\end{array}\right)~~~~~\mbox{and}~~~~\psi^+_j =
(c^+_{j\uparrow},c_{j_\downarrow}). 
$$
Using this in (1) we obtain 
\begin{equation}
 H = H^{\rm MF} + H' + H''\,,
\label{eqno4}
\end{equation}
where in site representation 
\begin{eqnarray}
 H^{\rm MF} & = & \sum_{ij} (r^2 t_{ij} - \mu\delta_{ij})\,
  \psi^+_i \hat\tau_3\psi_j + \sum_{ijs\alpha}(r^2 T'^\alpha_{ijs}
+ V^\alpha_{js}\delta_{ij}) u^\alpha_s \psi^+_i
\hat\tau_3\psi_j\nonumber \\
& + & \Lambda \sum_i (\psi^+_i \hat\tau_3\psi_i + r^2 - 1) +
   H^{\rm ph}\nonumber\,, \\
H' & = & \sum_{ij} [r \, t_{ij} \psi^+_i \hat\tau_3\psi_j +
   \Lambda\delta_{ij}] (\delta b_i + \delta b^+_j) + 
 \sum_{ijs\alpha} r \, T'^\alpha_{ijs}  u^\alpha_s \psi^+_i \hat\tau_3\psi_j
 (\delta b_i + \delta b^+_j)  \,,\\
H'' & = & \Lambda \sum_i \delta b^+_i  \delta b_i + \sum_{ij} \, 
  t_{ij} \psi^+_i \hat\tau_3\psi_j \delta b_i \delta b^+_j + 
  \sum_{ij\alpha s} T'^\alpha_{ijs} u^\alpha_s \psi^+_i
  \hat\tau_3\psi_j\, \delta b_i \delta b^+_j\,. \nonumber
\label{eqno5}
\end{eqnarray}
First term of $H'$ describes fermion--boson interaction, while
the last one fermion--phonon--boson interaction. More
complicated interactions are contained in $H''$. They will be
treated in a~mean field type of approximation. The important
point is that $H''$ significantly contributes to dynamics of
fluctuating boson field. In the mean field approximation for
electrons the boson fluctuations are described by effective
Hamiltonian 
\begin{equation}
 H_B = \sum_i (\Lambda\delta_{ij} + t_{ij} \langle c^+_{i\sigma}
  c_{j\sigma}\rangle) \delta b^+_j \delta b_i
\label{eqno6}
\end{equation}
For the sake of completeness we write down the Hamiltonian
for phonons which we assume to be described in the harmonic
approximation. In the second quantised form it reads
\begin{equation}
H_{\rm ph} = \sum_{q\nu} \hbar \omega_{q\nu}(a^+_{q\nu}a_{q\nu}
+  {1 \over 2})\,.
\label{eqno7}
\end{equation}

\bigskip\bigskip
%\noindent
\section{Equations of superconductivity}

\bigskip
To properly describe the superconducting state in the system at
hand one has to work in site representation. The important point
is that in considered $U = \infty$ limit the double
occupation of a~given site is strictly forbidden. This means
{\it inter alia} that correlation functions $\langle
c_{i\uparrow}c_{j\downarrow}\rangle$ describing superconducting
pairs vanish exactly for $i=j$, i.e. the on--site pairing is
forbidden. On the other hand the correlations of the type 
$\langle c^+_{i\sigma} c_{i\sigma}\rangle$ measure the average number 
of carriers at site $i$, and are allowed to enter into formula. 
This important fact has first been noted by Zieli/nski and 
coworkers \cite{Mierz,Ziel} and leads, as we shall see, to severe
changes in the form of Eliashberg equations.

To derive them we use the equation of motion method for the
double time thermodynamic Green's functions. For arbitrary
operators $A$ an $B$ it reads (we employ here the Zubariev
notation \cite{Zubariev})
\begin{eqnarray*}
 \omega \ll A\vert B\gg_\omega & = & \langle [A,B]_\pm \rangle +
    \ll[A,H]_- \vert B\gg_\omega \\
& = & \langle[A,B]_\pm\rangle - \ll A\vert[B,H]_-\gg_\omega\,, 
\end{eqnarray*}
where $\ll A\vert B\gg_\omega$ is the Fourier transformed,
frequency dependent, retarded Green's function. $[A,B]_\pm$
denotes anticommutator ($+$) or commutator ($-$).

Standard procedure \cite{KIWKUZ} leads to the following equation
for the matrix Green's function (GF)
\begin{equation}
\sum_l \{[\omega\hat\tau_0 + (\mu - \Lambda)\hat\tau_3] \, \delta_{il} 
 - (r^2 + \langle\delta b_i\delta b^+_l\rangle)\,t_{il} \hat\tau_3
 - \widehat M_{il}(\omega)\} \ll \psi_l\vert \psi^+_j\gg_\omega =
 \delta_{ij} \hat\tau_0 .
\label{eqno8}
\end{equation}
Here $\hat\tau_0 = \left(\begin{array}{ll} 1 & 0 \\ 0 &
1\end{array}\right),~~~\hat\tau_1 = \left(\begin{array}{ll} 0 & 1 \\ 1 &
0\end{array}\right),~~~\hat\tau_3 = \left(\begin{array}{ll} 1 & ~0 \\ 0 &
-1\end{array}\right)$ are Pauli matrices and $\widehat M_{il}$
denotes the matrix self--energy. Due to complicated interactions
in the Hamiltonian $\widehat M_{il}$  contains a~number of terms. Here we write
down few most important contributions.

The contribution from fermion--phonon scattering is given by
\begin{equation}
\widehat M^{\rm ph}_{ij}(\omega) = \sum_{i'j'}
 \sum_{\alpha\alpha'ss'} (r^2T^\alpha_{ij's} + V^\alpha_{j's}
 \delta_{ij'}) (r^2 T^{\alpha'}_{i'js'} + V^{\alpha'}_{js'} \delta_{i'j})
 \hat\tau_3 \ll u^\alpha_s \psi_{j'} \vert u^{\alpha'}_{s'}
 \psi^+_{i'} \gg_\omega \, \hat\tau_3.
\label{eqno11}
\end{equation}

The contribution from linear boson--fermion scattering reads
\begin{equation}
 \widehat M^B_{ij}(\omega) = \sum_{i'j'} r^2 t_{ij'}t_{i'j} \hat\tau_3 
\ll\phi^B_{ij'} \psi_{j'} \vert\phi^B_{i'j}
\psi^+_{i'}\gg_\omega \hat\tau_3,
\label{eqno9}
\end{equation}
with $\phi^B_{ij} = \delta b_i + \delta b^+_j$. 

Related, but probably less important contribution from quadratic
boson--fermion scattering (from $H''$)
\begin{equation}
\widehat M'^B_{ij}(\omega) = \sum_{i'j'} t_{ij'}t_{i'j} \hat\tau_3 
 \ll \delta b_i \delta b^+_{j'}\psi_{j'} \vert \delta b_{i'}
 \delta b^+_j \psi^+_{i'} \gg_\omega \hat\tau_3.
\label{eqno10}
\end{equation}
This contribution is absent if one approximates $H''$ in mean
field like manner.

There are also two contributions from three particle scattering
events: fermion--phonon--slave boson. From $H'$ part of the
Hamiltonian we get
\begin{equation}
\widehat M^{ph-B}_{ij}(\omega) = \sum_{i'j'} \sum_{\alpha\alpha'ss'}
r^2 T^\alpha_{ij's} T^{\alpha'}_{i'js'} \hat\tau_3 
 \ll\phi^B_{ij'} u^\alpha_s \psi_{j'} \vert\phi^B_{i'j} u^{\alpha'}_{s'}
 \psi^+_{i'}\gg_\omega \hat\tau_3,
\label{eqno12}
\end{equation}
while from second term in $H''$
$$
 \widehat M'^{ph-B}_{ij}(\omega) = \sum_{i'j'} \sum_{\alpha\alpha'ss'}
 T'^\alpha_{ij's} T'^{\alpha'}_{i'js'} \hat\tau_3 
 \ll u^\alpha_s \delta b_i \delta b^+_{j'} \psi_{j'} \vert u^{\alpha'}_{s'}
 \delta b_{i'} \delta b^+_j \psi^+_{i'}\gg_\omega \hat\tau_3.
$$
To get real frequency axis Eliashberg equations in a~standard
form \cite{Wons} one has to express the higher order Green's
function appearing on the rhs' of various selfenergy pieces by
the GF's of fermions $\ll \psi_{\vec k}\vert\psi^+_{\vec
k}\gg_\omega$, bosons $\ll \delta b_{\vec p}\vert \delta
b^+_{\vec p}\gg_\omega$ and ,,phonons'' $\ll \phi_{\vec q\nu}
\vert \phi_{-\vec q\nu}\gg_\omega$. This is easily achieved with help
of spectral representation and decoupling  of various time
correlation functions as e.g.
$$
 \langle \delta b_{\vec p}(t) \phi_{q\nu}(t) \psi_{k'}(t) 
\delta b^+_p(0)  \phi_{q\nu}(0) \psi^+_{k'}(0)\rangle
\approx \langle \delta b_p(t) \delta b^+_p(0) \rangle 
 \langle \phi_{q\nu}(t) \phi_{-q\nu}(0)\rangle 
 \langle \psi_{k'}(t) \psi^+_{k'}(0)\rangle. 
$$
This approximation neglets vertex corrections. In the present
situations the vertex corrections stem from fermion--phonon,
fermion--boson and also more complicated interactions. As
already mentioned the assumption of small electron--phonon
coupling allows us to neglect fermion--phonon vertex
corrections. The same is not true for the fermion--boson
vertices. They do not contain small parameters and should be
taken into account. Their importance has also been stressed in
previous studies \cite{Grilli,Zeyher}. 
We shall postpone discussion of vertex corrections and
concentrate on the anisotropy of Eliashberg equations.

The next step is Fourier transform of the various selfenergy
parts. Defining the Fourier components of $\widehat M_{ij}$ by 
\begin{equation}  
 \hat\Sigma_{\vec k}(\omega) = {1 \over N} \sum_{ij} \,
 e^{-i\vec k(\vec R_i-\vec R_j)} \, \widehat M_{ij}(\omega),
\label{eqno13}
\end{equation}
we get
\begin{eqnarray}
\hat\Sigma^{\rm ph}_{\vec k}(\omega) & = & {1 \over N^2} 
 \sum_{\vec k'\vec q\nu} \vert M^\nu_{\vec k,\vec q}\vert^2\hat\tau_3
 \ll \phi_{\vec q\nu} \psi_{\vec k'}\vert \phi_{-\vec q\nu}
\psi^+_{\vec k'} \gg_\omega \hat\tau_3 \, {1\over N} \sum_{i'j'} \,
 e^{-i(\vec k'+\vec q-\vec k)(\vec R_{i'}-\vec R_{j'})},\nonumber\\
\nonumber \\
\hat\Sigma^B_{\vec k}(\omega) & = & {1 \over N^2} 
 \sum_{\vec k',\vec q} \left\{\epsilon^2_{\vec k-\vec q} \hat\tau_3 
 \ll \delta b_{\vec q} \psi_{\vec k'} \vert \delta b^+_{\vec q}
 \psi^+_{\vec k'}\gg_\omega \hat\tau_3 \, {1\over N} \sum_{i'j'}
 e^{i(\vec k'+\vec q-\vec k)(\vec R_{j'}-\vec R_{i'})} \right. \nonumber \\
 & + & \left. \epsilon^2_{\vec k} \, \hat\tau_3
 \ll \delta b^+_{\vec q} \psi_{\vec k'} \vert \delta b_{\vec q}
 \psi^+_{\vec k'} \gg_\omega \hat\tau_3 \, {1\over N} \sum_{i'j'}
 e^{i(\vec k'-\vec k-\vec q)(\vec R_{j'}-\vec R_{i'})} \right\},\nonumber \\
\nonumber \\
\hat\Sigma^{\rm ph-B}_{\vec k} & = & {1\over N^3} \sum_{\vec k'\vec q,\vec p}
 \left\{r^2 \vert \tilde M^\nu_{\vec k,\vec q,\vec p} \vert^2
 \hat\tau_3 \ll \delta b_{\vec p} \phi_{\vec q\nu} \psi_{\vec k'}
 \vert \delta b^+_{\vec p} \phi_{-\vec q\nu} \psi^+_{\vec k'}\gg_\omega 
 \hat\tau_3 \right. \nonumber \\
&&  {1\over N} \sum_{i'j'} e^{i(\vec k-\vec q)(\vec R_{i'}- 
 \vec R_{j'})} + r^2  \vert \tilde M^\nu_{\vec k,\vec q,\vec p=0}\vert^2
 \hat\tau_3 \ll \delta b^+_{\vec p} \phi_{\vec q\nu} \psi_{\vec k'}
\vert \delta b_{\vec p} \phi_{-\vec q\nu} \psi^+_{\vec k'}\gg_\omega 
 \hat\tau_3 \,\nonumber \\
&& \left. {1\over N} \sum_{i'j'} \, e^{i(\vec k+\vec p-\vec q)(\vec
 R_{i'}-\vec R_{j'})} \right\}.
\label{eqno14}
\end{eqnarray}
Here $\phi_{q\nu} = a_{q\nu} + a^+_{-q\nu}$ denotes the phonon
field, $a^+_{q\nu}$ is creation operator of the $q\nu$ phonon.

Electron--phonon matrix elements $M^\nu_{kq}$ and $\tilde
M^\nu_{kqp}$ are defined as
\begin{eqnarray}
&& \vert M^\nu_{\vec k,\vec q}\vert^2 = \vert g_\nu(\vec q) \vert^2  \,
 \vert \sum_\alpha \, e^\alpha_\nu(\vec q) [r^2 V_{c}(\tilde 
  v^\alpha_{\vec k-\vec q} - \tilde v^\alpha_{\vec k})
  + V_{i} V^\alpha_{\vec q}] \vert^2, \nonumber \\
\\
&& \vert\tilde M^\nu_{\vec k,\vec q,\vec p}\vert^2 = 
V^{2}_{c} r^2 \vert g_\nu(\vec q)\vert^2 \, \vert \sum_{\alpha}
e^\alpha_\nu(\vec q) 
  (\tilde v^\alpha_{\vec k-\vec q-\vec p} - \tilde v^\alpha_{\vec
   k-\vec p})\vert^2,\nonumber
\label{eqno15}
\end{eqnarray}
and $g_\nu(\vec q) = (\hbar//2\omega_\nu(\vec q) \cdot M)^{1//2}$, $\tilde
v^\alpha_k = {1\over \hbar} {\partial\epsilon_{\vec k} \over 
\partial k^\alpha}$,~~$e^\alpha_\nu(\vec q)$ is the $\alpha$--th
component of the phonon polarisation operator, while
$V^\alpha_{\vec q}$ is the Fourier transform of the ionic part of
electron--phonon interaction, $\omega_\nu(\vec q)$ denotes the phonon
dispersion and $M$ --- the ionic mass. $V_{c}$ ($V_{i}$) denotes the
strength of the covalent (ionic) part of electron -- phonon interaction.

The factors  of the type 
$$
 {1 \over N} \sum_{i'j'} \, e^{i\vec k(\vec R_{i'}-\vec R_{j'})}
$$
are to be treated very carefully. As already mentioned the sum
goes over all sites $i'$ and $j'$ if we evaluate the normal
contributions to the self--energy matrix i.e. components (1,1) and
(2,2) of the self--energy matrix  which are proportional to the GF of the type
$\ll\hat O c_{i'\sigma}\vert \hat O'c^+_{j'\sigma}\gg$ 
($\hat O, \hat O'$ are arbitrary operators). In such
a~case the above sum reduces to $N\delta_{\vec k,0}$. When
calculating off--diagonal elements of the selfenergy which
depends on the GF's like $\ll \hat O c_{i'\uparrow}\vert \hat
O'c_{j'\downarrow}\gg$ it is reasonable to assume $i'$ and $j'$
to be nearest neighbours, as in the $U = \infty$ limit only
nearest neigbhour pairs will probably survive. In this situation
the sum reduces to $\gamma(\vec k) = -\epsilon(\vec k)//2t$. 

The approximation of the sum to $(i',j')$ being nearest
neighbour sites relies on the experimental fact that the
superconducting coherence length is small: of order of
lattice spacing. In an effective theory like the one presented
here it translates to nearest neighbour pairs.

To proceed we expand matrix selfenergy $\hat\Sigma_k(\omega)$ as
\begin{equation}
 \hat\Sigma_k(\omega) = \omega[1 - Z_k(\omega)] \hat\tau_0 + 
 \phi_k(\omega) \hat\tau_1 + \chi_k(\omega)\hat\tau_3
\label{eqno16}
\end{equation}
and write down the equations for various parts of it. We get
\begin{eqnarray}
\omega[1 - Z^{\rm ph}_{\vec k}(\omega)] & = & 
 {1\over 2} \int d\omega_1 \int d\omega_2 
 {th {\beta\omega_1\over 2} + cth {\beta\omega_2 \over 2} \over 
 \omega - \omega_1 - \omega_2} \, {1\over N} \sum_{\vec k'} \,
 K^{\rm ph}_{\vec k\vec k'}(\omega_2) \cdot \nonumber \\
&& (-{1\over \pi}) \, Im {\omega_1 Z_{\vec k'}(\omega_1) \over 
   D_{\vec k'}(\omega_1)} \nonumber \\
\nonumber \\
\phi^{\rm ph}_{\vec k}(\omega) & = & 
 {1\over 2} \int d\omega_1 \int d\omega_2 
 {th {\beta\omega_1\over 2} + cth {\beta\omega_2 \over 2} \over 
 \omega - \omega_1 - \omega_2} \, {1\over N} \sum_{\vec k'} \,
 \tilde K^{\rm ph}_{\vec k\vec k'}(\omega_2) \cdot \nonumber \\
&&  (-{1\over \pi}) \, Im {-\phi_{\vec k'}(\omega_1) \over 
   D_{\vec k'}(\omega_1)} \nonumber \\
\nonumber \\
\chi^{\rm ph}_{\vec k}(\omega) & = & 
 {1\over 2} \int d\omega_1 \int d\omega_2 
 {th {\beta\omega_1\over 2} + cth {\beta\omega_2 \over 2} \over 
 \omega - \omega_1 - \omega_2} \, {1\over N} \sum_{\vec k'} \,
 K^{\rm ph}_{\vec k\vec k'}(\omega_2) \cdot \nonumber \\
&& (-{1\over \pi}) \, Im {r^2 \epsilon_{\vec k'} - \mu + \Lambda
+ \sum_{\vec q} \epsilon_{\vec k'-\vec q} \langle \delta
b_{\vec q} \delta b^+_{\vec q}\rangle + \chi_{\vec k'}(\omega_1)
\over D_{\vec k'}(\omega_1)}
\label{eqno17}
\end{eqnarray}
where we denoted $D_{\vec k'}(\omega_1) =
[\omega_1 Z_{\vec k'}(\omega_1)]^2 - [\phi_{\vec k'}(\omega_1)]^2 - 
[r^2 \epsilon_{\vec k'} - \mu + \Lambda + \sum_{\vec q}
\epsilon_{\vec k'-\vec q} \langle \delta b_{\vec q} \delta 
b^+_{\vec q}\rangle + \chi_{\vec k'}(\omega_1)]^2$ and 
\begin{eqnarray}
&& K^{\rm ph}_{\vec k\vec k'}(\omega_2) = \sum_\nu 
\vert M^\nu_{\vec k,\vec k-\vec k'}\vert^2 (-{1\over\pi}) 
Im \ll \phi_{\vec k-\vec k',\nu} \vert\phi_{-\vec k+\vec k',\nu}
\gg_{\omega_2+io}, \\
\label{eqno18}
&& \tilde K^{\rm ph}_{\vec k,\vec k'}(\omega_2) = {1\over N} \sum_{\vec q\nu} 
 \, \vert M^\nu_{\vec k,\vec q}\vert^2 (-{1\over \pi})\,
Im \ll \phi_{\vec q,\nu} \vert\phi_{-\vec q\nu}\gg_{\omega_2+io} \,
 \gamma(\vec k-\vec q-\vec k').
\label{eqno19}
\end{eqnarray}
Completely analogous set of equations is obtained for 
contributions to $\hat\Sigma^B_{\vec k}(\omega)$. The only
difference is that $K^{\rm ph}_{\vec k,\vec k'}$ and $\tilde
K^{\rm ph}_{\vec k,\vec k'}$ will be replaced by $K^B_{\vec
k,\vec k'}$ and $\tilde K^B_{\vec k,k'}$ given by
\begin{equation}  
 K^B_{\vec k\vec k'}(\omega_2) = \epsilon^2_{\vec k'} (-{1\over \pi}) \, 
Im \ll\delta b_{\vec k-\vec k'} \vert \delta b^+_{\vec k-\vec
k'}\gg_{\omega_2 +io} + \epsilon^2_{\vec k}  (-{1\over \pi}) \, 
Im \ll\delta b_{-\vec k+\vec k'} \vert \delta 
b_{-\vec k+\vec k'}\gg_{\omega_2 +io},
\label{eqno20}
\end{equation}
and 
\begin{eqnarray}
 \tilde K^B_{\vec k,\vec k'}(\omega_2) & = & {1\over N} \sum_q \,
  \epsilon^2_{\vec k-\vec q} (-{1\over \pi}) \, 
 Im \ll\delta b_{\vec q} \vert \delta b^+_{\vec q} \gg_{\omega_2 +io} \,
 \gamma(\vec k - \vec q - \vec k')\nonumber \\
&& +  \epsilon^2_{\vec k} {1\over N}  \,  \sum_q \left(-{1\over\pi}\right)\,
  Im \ll\delta b^+_{\vec q} \vert \delta b_{\vec q} \gg_{\omega_2 +io} \,
 \gamma(\vec k + \vec q - \vec k').
\label{eqno21}
\end{eqnarray}
Due to complicated structure of the GF entering the expression
for $\tilde M^{ph-B}_{\vec k}(\omega)$ the corresponding
Eliashberg equations on the real frequency axis have more
complicated form
\begin{eqnarray}
&& \omega[1 - Z^{ph-B}_{\vec k}(\omega)] =
 {1\over 4} \int d\omega_1 \int d\omega_2 \int d\omega_3
 {th {\beta\omega_1\over 2}  (cth {\beta\omega_2\over 2} + 
 cth {\beta\omega_3\over 2}) + cth {\beta\omega_2\over 2} 
 cth {\beta\omega_3\over 2} + 1 \over 
 \omega - \omega_1 - \omega_2 -\omega_3}  \nonumber\\
&& \hspace{2cm}  {1\over N} \sum_{\vec k'} \, K^{ph-B}_{\vec k\vec k'} 
(\omega_2,\omega_3) (-{1\over \pi}) \, Im {\omega_1 Z_{\vec k'}(\omega_1) 
\over D_{\vec k'}(\omega_1)} \nonumber\\
\nonumber \\
&& \phi^{ph-B}_{\vec k}(\omega) = 
 {1\over 4} \int d\omega_1 \int d\omega_2 \int d\omega_3
 {th {\beta\omega_1\over 2} (cth {\beta\omega_2\over 2} + 
 cth {\beta\omega_3\over 2}) + cth {\beta\omega_2\over 2} 
 cth {\beta\omega_3\over 2} + 1 \over 
 \omega - \omega_1 - \omega_2 -\omega_3}  \,\nonumber\\
&& \hspace{2cm}  {1\over N} \sum_{\vec k'} \, 
  \tilde K^{ph-B}_{\vec k\vec k'} 
 (\omega_2,\omega_3) (-{1\over \pi}) \, Im {-\phi_{\vec k'}(\omega_1) 
 \over D_{\vec k'}(\omega_1)} \nonumber\\
\nonumber\\
&& \chi^{ph-B}_{\vec k}(\omega) =
  {1\over 4} \int d\omega_1 \int d\omega_2 \int d\omega_3
 {th {\beta\omega_1\over 2}  (cth {\beta\omega_2\over 2} + 
 cth {\beta\omega_3\over 2}) + cth {\beta\omega_2\over 2} 
 cth {\beta\omega_3\over 2} + 1 \over 
 \omega - \omega_1 - \omega_2 -\omega_3}  \nonumber\\
&& \hspace{2cm}  {1\over N} \sum_{\vec k'} \, K^{ph-B}_{\vec k\vec k'} 
  (\omega_2,\omega_3) (-{1\over \pi}) \, Im {r^2\epsilon_{\vec k'} 
  - \mu + \Lambda + \sum_{\vec q} \epsilon_{\vec k'-\vec q} \langle \delta
  b_{\vec q} \delta b^+_{\vec q}\rangle + \chi_k'(\omega_1) 
  \over D_{k'}(\omega_1)},
\label{eqno22}
\end{eqnarray}
with 
\begin{eqnarray}
&& K^{ph-B}_{\vec k\vec k'}(\omega_2,\omega_3) = 
{1\over N} \sum_{\vec p\nu} \left\{r^2 \vert \tilde M^\nu_{\vec k, 
 \vec q,\vec p} \vert^2 (-{1\over\pi}) \, Im
 \ll \phi_{\vec k\nu} \vert \phi_{-\vec k\nu} 
  \gg_{\omega_2+io} (-{1\over \pi}) Im 
\ll \delta b_{\vec p} \vert \delta b^+_{\vec p}
  \gg_{\omega_3+io} \right. \nonumber \\
&&\hspace{1.5cm}  \left. + r^2 \vert \tilde M^\nu_{\vec k, 
 \vec k+\vec p,\vec p,0} \vert^2 (-{1\over\pi}) \, Im
 \ll \phi_{\vec k+\vec p\nu} \vert \phi_{-\vec k-\vec p,\nu} 
  \gg_{\omega_2+io} (-{1\over \pi}) \, Im
\ll \delta b^+_{\vec p} \vert \delta b_{\vec p} \gg_{\omega_3+io} \right\} 
\label{eqno23}
\end{eqnarray}
and
\begin{eqnarray}
&& \tilde K^{ph-B}_{\vec k\vec k'}(\omega_2,\omega_3) =
 {1 \over N^2} \sum_{\vec q\vec p} \left\{r^2 \vert \tilde M^\nu_{\vec k, 
 \vec q,\vec p} \vert^2 (-{1\over\pi}) \, Im
 \ll \phi_{\vec q\nu} \vert \phi_{-\vec q\nu} \gg_{\omega_2+io} 
 (-{1\over \pi}) \, Im \ll \delta b_{\vec p} \vert \delta b^+_{\vec p}
 \gg_{\omega_3+io} \cdot \right.   \nonumber\\
&& \left. \cdot \gamma(\vec k -\vec q) + r^2 \vert \tilde M^\nu_{\vec
k,\vec q,\vec p=0} \vert^2 (-{1\over \pi}) \, Im
 \ll \phi_{\vec q\nu} \vert \phi_{-\vec q\nu} \gg_{\omega_2+io} 
 (-{1\over \pi}) \, Im \ll \delta b^+_{\vec p}  \vert \delta b_{\vec p}   
\gg_{\omega_3+io} \, \gamma(\vec k+\vec p -\vec q)
\right\}\nonumber\\ 
\label{eqno24}
\end{eqnarray}
Equations (17--24) supplemented with expressions for phonon and slave --
boson Greens functions form a~complete set and have to be solved
selfconsistently. 

There are few differences between them and
usual Eliashberg equations for superconductors. The most
important is the presence of two different types of kernels
$K_{kk'}$ and $\tilde K_{kk'}$ determining normal
($Z_{k'},\chi_{k'}$) and anomalous ($\phi_{k'}$) parts of the
self--energy, respectively.

Other point worth to mention is strong renormalisation of the
bare electron spectrum $\epsilon_{\vec k} = -2t\gamma(\vec k)$.
At the mean field level of the theory it is both shifted and mass
renormalised to $r^2\epsilon_{\vec k} + \Lambda$. Beyond mean field
level for bosons the nontrivial energy dependence comes from
renormalisation of the fermion (matrix) Greens function given by the 
$\chi^B_{\vec k}(\omega)$ factor. Note, that from the equation for 
$\chi^B_{\vec k}(\omega)$ and definition (20) of $K^B_{\vec
k\vec k'}(\omega)$ we have
\begin{equation}
 \chi^B_ {\vec k}(\omega) = C(\omega) + \epsilon^2_{\vec k} \,C_1(\omega),
\label{eqno25}
\end{equation}
where $C(\omega)$ and $C_1(\omega)$ are weakly $\vec k$
dependent functions to be calculated from
equation (17). To take this effect into account one has to
solve the equations at each point of the Brillouin zone --- a very
difficult task.

\bigskip\bigskip
\noindent
\section{Results and discussion}

\bigskip
The complete solution of Eliashberg equations which are integral
equations with complicated kernels would require finding all
functions $Z_k(\omega)$, $\phi_k(\omega)$, $\chi_k(\omega)$ at
all frequencies and all points in the Brillouin zone. In the
standard approach the valid argument was that most important changes
are to be expected at the Fermi level. The various pieces of
selfenergies were thus averaged over the wave vectors lying on
the Fermi surface. In this way it was possible to reduce the
equations to the integral equation of the single variable: the 
frequency.
It is not obvious that the same arguments do apply in the
present case especially to the boson contribution to
self--energy. 

Sticking to the mean field approximation for slave bosons leads
to $\delta b_q = \delta b^+_q \equiv 0$. In this approximation only phonon
contribution to $\Sigma_{\vec k}$ survives,
$\hat\Sigma_k(\omega) = \hat\Sigma^{\rm
ph}_k(\omega)$. In this case most of arguments developed
previously \cite{Wons} apply and one can use known parametrisations of
solutions to Eliashberg equations. The relevant parameters
describing superconducting transition temperature $T_c$ are the
typical phonon frequency $\omega_D$ and electron--phonon
parameter $\lambda_{\rm e-ph}$. The last parameter is determined
by the kernel $K^{\rm ph}$. In our theory we do
have two different kernels as also found previously by Zieli/nski
and coworkers \cite{Mierz} \cite{Ziel}.

We thus define two coupling constants $\lambda_{\rm e-ph}$ and
$\tilde\lambda_{\rm e-ph}$. The first one is related to wave
function renormalisation and other describes genuine
superconducting coupling in strongly correlated system. 
We define $\lambda_{\rm el-ph}$ by \cite{Wons}
\begin{equation}
\lambda_{\rm e-ph} = 2\int {d\omega \over \omega} \,
\alpha^2F(\omega) 
\label{eqno26}
\end{equation}
where
$$
\alpha^2F(\omega) = {1\over N} \sum_{\vec k \vec k'} K^{\rm ph}_{\vec k
\vec k'}(\omega) 
\, \delta(r^2\epsilon_{\vec k} - \mu + \Lambda) \delta(r^2\epsilon_{\vec
k'} - \mu
+ \Lambda) \,// 
 \sum_{\vec k} \delta(r^2\epsilon_{\vec k} - \mu + \Lambda) 
$$
Similar expression defines $\tilde\lambda_{\rm e-ph}$ except
that $\tilde K^{\rm ph}_{kk'}(\omega)$ enters. In the general
case there exist four more parameters: $\lambda_{e-B}$,
$\tilde\lambda_{e-B}$, $\lambda_{e-ph-B}$ and
$\tilde\lambda_{e-ph-B}$. 

For numerical purposes it is convenient to take optical
phonons with $\omega_{q\nu} = \omega_o$. In this case the
frequency integral in (26) can be calculated very
easily. 

Strong wave vector dependence of the kernels $\tilde
K^{ph}_{\vec k\vec k'}$ makes the question of symmetry of the order
parameter very important. To get some information on this point
we calculate various symmetry components $\lambda^i_{e-ph}$ of
the pairing interaction. We define them as
$$
 \tilde\lambda^i_{e-ph} = {1\over N} \sum_{\vec k\vec k'} \,
 \tilde K^{ph}_{\vec k\vec k'} \delta (r^2\epsilon_{\vec k} - \mu +
 \lambda) \, \delta(r^2\epsilon_{\vec k'} - \mu + \lambda) 
  g_i(\vec k) g_i(\vec k')//A_i
$$
We have taken $g_1 = \cos k_x a + \cos k_y a$, $g_2 = \cos k_x a
- \cos k_y a$, $g_3 = \sin k_x a +  \sin k_y a$ and $g_4 = \sin k_x
a- \sin k_y a$; $A_i = \sum_{\vec k} g^2_i(\vec k) \delta (r^2\epsilon_{\vec
k} - \mu + \lambda)$. 

There are few parameters which control the behaviour of the system.
These are: carrier concentration $n$, strength of the electron --
phonon interaction parameters $V_c$ and $V_i$, the frequency $\omega_0$
of the
optical phonon etc. All the energies are measured in units of $D=2t$,
and for numerical purposes we have taken $\omega_0 = 0.02$, and the
value of the ionic mass 
$M$ appropriate for oxygen.

In this work we shall concentrate on the results obtained from the
kernel $\tilde K^{\rm ph}_{\vec k \vec k\prime}$, describing anomalous
part of the self--energy. In the previous work\cite{KIW} we have
analysed in some detail the properties of the system resulting from the
kernel entering normal self--energy.

We start the discussion by showing the results
obtained for $V_{i}=0$. The covalent part of the interaction is
strongly $\vec k$ dependent. It is described by the matrix element
$\vert M_{\vec k \vec q} \vert^2$ in equation (15). In figure (1a) we
show the results obtained by making the same approximation as in previous
treatment of the similar model,\cite{Ziel} where $\vert M_{\vec k \vec
q} \vert^2$   has been replaced by wave
vector independent constant multiplied by $r^4$. The concentration
dependence of {\it s, p} and $d$
components of the coupling constant are shown by solid, dotted and
dashed lines, respectively.  It is the factor $r^4 = (1-n)^2$, which
strongly supresses the values of $\tilde\lambda_i$ at higher electron
concentrations. The important point is that for small doping of holes
the d-wave component is the largest one. This signals the possibility
of the d-wave pairing in this parameter range. For $\delta=1-n > 0.3$
the p-wave pairing takes over. At quite small electron concentrations
extended s-wave seem to be most stable pairing.

In figure (1b) we show similar set of data obtained with $\vert M_{\vec
k \vec q} \vert^2$ replaced by its Fermi surface average $M^2_F$. 
Due to oscilatory dependence of $\vert M_{\vec k \vec q} \vert$ on wave
vectors the average value of it is strongly concentration dependent (solid
line in figure (1b)) and
reduced in comparison to 1 (the value arbitrarily assumed for data
shown in Fig.(1a)).
Small values of $M_F$ at large $n$ are due again to $r^4$ factor, while
at small $n$ to phase space restrictions.

Taking the full wave vector dependence of $\vert M_{\vec k \vec q}
\vert$ leads to the results shown in figure (1c). Solid line in this
figure shows total coupling $\tilde\lambda_{\rm el-ph}$ calculated from (26).
Other lines represent the symmetry components $\tilde\lambda_i$ as indicated.
The values of all couplings are very small. Thus in the mean field
approximation for slave bosons one is not able to get realistic values
of the superconducting transition temperature from covalent part of the
electron-phonon
interaction.  Let us look at the relative values of the various couplings.
At large electron concentration (small hole dopping) the d- wave
component of $\tilde\lambda$ exceeds by far all other components. 

Bigger values of coupling constants are obtained when ionic part of the
electron -- phonon interaction is taken into account. Figure (2a) shows
the concentration dependence of $\tilde\lambda$ (solid line) and
$\tilde\lambda_i$ 
(dashed lines) obtained for $V_c=0$ and $V_i=1$ (in units of $D$). The
values of the coupling parameters are quite realistic now. The nice feature of
this result is the possibility of d symmetry pairing at small hole
concentration. The presence of both ionic and covalent parts of
electron-phonon interaction leads to small changes of the phase
diagram (cf. Fig.2(b)) . The d symmetry remains most probable at small
hole dopings 
$\delta < 0.3$. For intermediate carrier concentrations p-wave pairing
is most stable, while extended s-wave is stable at small electron
concentration $n < 0.3$. The slight decrease of the values of couplings
as compared to figure (2a) is due to interference effects between
interaction channels (cf.
expression (15) for $\vert M_{\vec k \vec q} \vert^2$).
%%%%%%%%%%%%%%%added material%%%%%%%%%%%%%%%%%%%%%%%%%%%%%%%%%%%%%%%%%
In this work we have concentrated on the calculation of the electron --
phonon coupling constants. To obtain realistic estimations of the
concentration dependence of the superconducting transition temperature
one has to take simultaneously into account the condensation of the
(slave) bosons. 
Boson condensation temperature $T^B_c$ of the 3d massive bosons is
proportional to the $2/3$ power of the boson concentration $(1-n)$. For
the masless bosons the power changes to $1/3$ in 3d case and $1/2$ for
interesting case of two dimensional system. Becouse superconducting
transition temperature, in the systems with covalent coupling (c.f.
figures 1a,1b,1c), also vanishes for $n \rightarrow 1$, the additional
condition for Bose -- Einstein condensation can  change the phase
diagram in quantitative way only. The situation is different in systems
with ionic or mixed (covalent plus ionic) interactions and especially
for the phase with $d$--wave symmetry of the order parameter. The
coupling constant $\lambda_d$ takes on sizeable values (c.f. figs. 2a
and 2b) at low hole concentration. In this case the additional
condition will modify the $T_c vs. n$ phase diagram making it
qualitatively more similar to that obtained for covalent electron --
phonon interaction.
%%%%%%%%%%%%%%%%%%%%%%%%%%%%%%%%%%%%%%%%%%%%%%%%%%%%%%%%%%%%%%%%%%%%%

In conclusion we have presented the derivation of the Eliashberg
equations for the system with strong electron -- electron and
electron -- phonon interactions. The results show the importance
of the correlations which make the superconducting order
parameter very anisotropic. Its symmetry depends on
the carrier concentration. We have illustrated the theory by showing the
results of calculations of coupling constants in the mean field
approximation for slave bosons. 

As already mentioned beyond mean field
level there are important changes of the structure of the theory. The
mixed fermion -- phonon -- slave boson contribution to the selfenergy
and corresponding coupling constants will overtake the fermion --
phonon contributions considered here. The $\chi$ component of the
selfenergy due to fermion -- slave boson scattering acquires nontrivial
energy $\epsilon_{\vec k}$ dependence. The study of both these effects
will be reported in the forthcomming paper.

\bigskip\bigskip
\noindent
{\bf Acknowledgements:} This work has been supported by KNB grant
No. 2P 302 070 06. Stimulated discussions with profs. M.~Grilli,
J.~Ranninger, R.~Zeyher and J.~Zieli\'n done during
acknowledged. Part of the work has been done during
author's stay in ISI Torino sponsored by the European Community.

\begin{figure}
\caption{The concentration dependence of the coupling constants
$\tilde\lambda_i$ calculated for a model with $V_i=0, V_c=1$ and: (a)
constant value of the 
electron -- phonon matrix element $\vert M_{\vec k \vec q} \vert=1.0$,
(b) $\vert M_{\vec k \vec q} \vert$ averaged over the Fermi surface and
(c) the actual value $\vert M_{\vec k \vec q} \vert$ with full wave
vector dependence taken into account}
\end{figure}
\begin{figure}
\caption{The pairing coupling constants {\it vs} $n$ for a model with:
(a) ionic 
electron -- phonon interaction ($V_i=1, V_c=0$) and (b) both ionic and
covalent interactions ($V_i=1, V_c=1$)}
\end{figure}
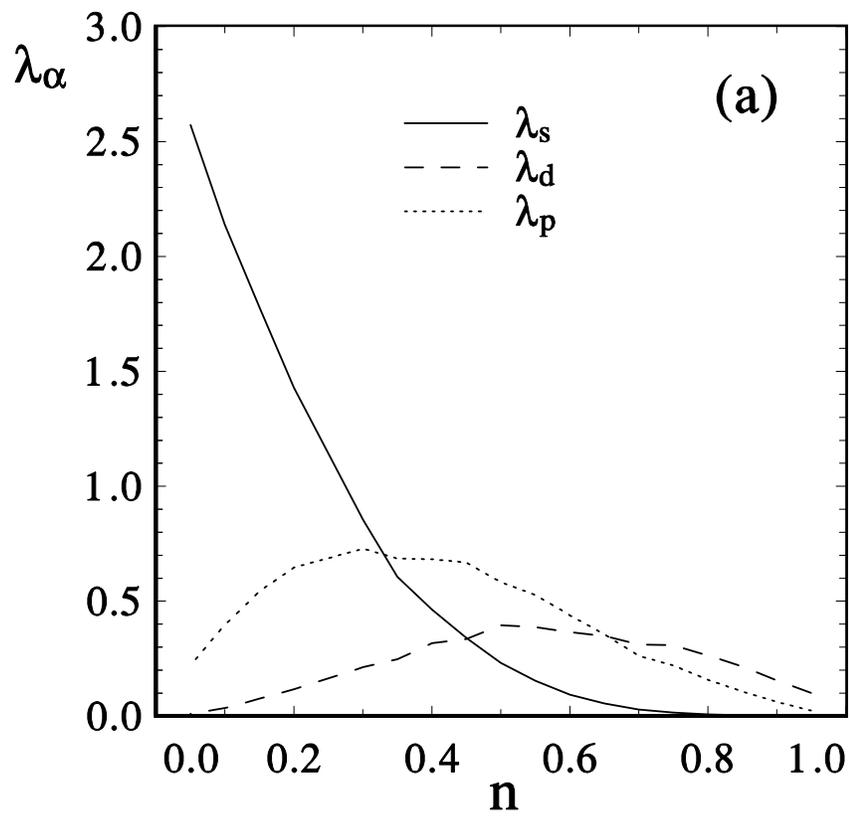
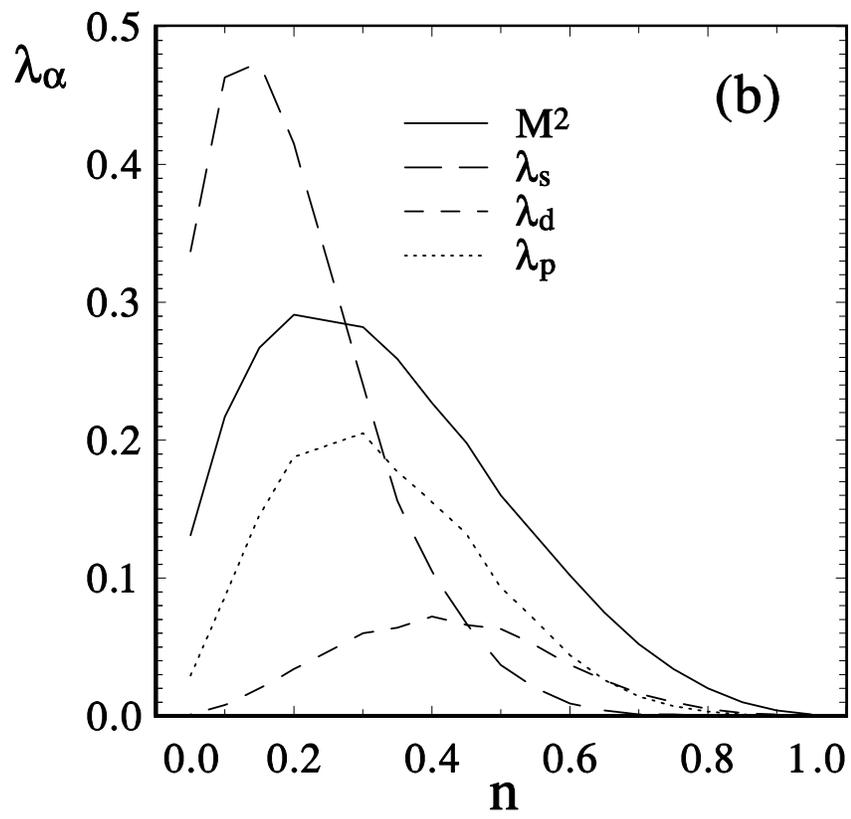
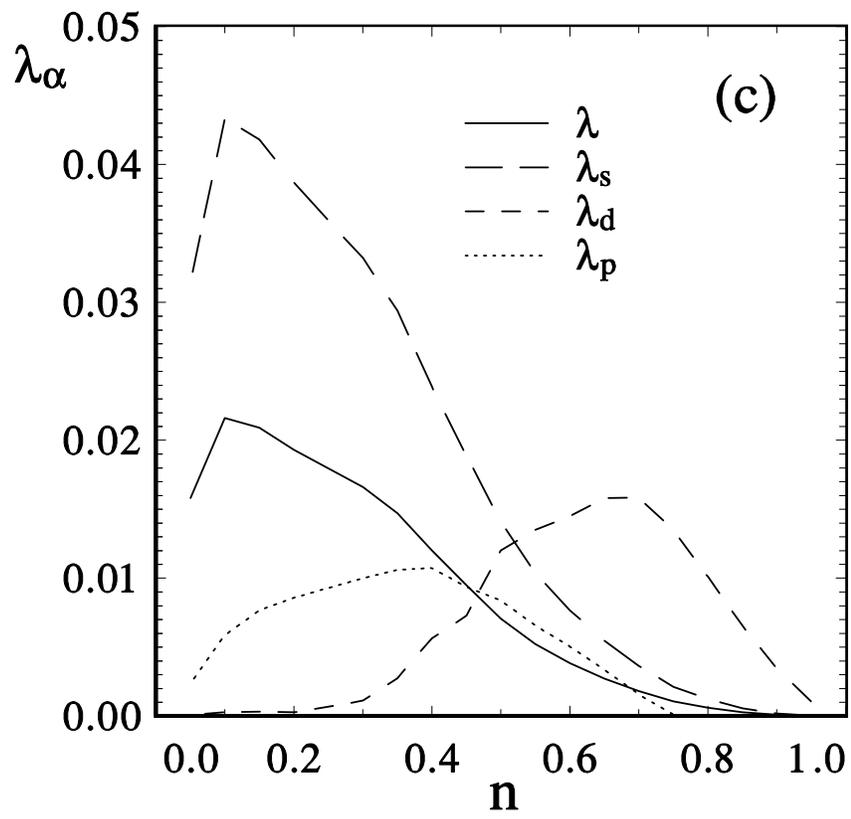
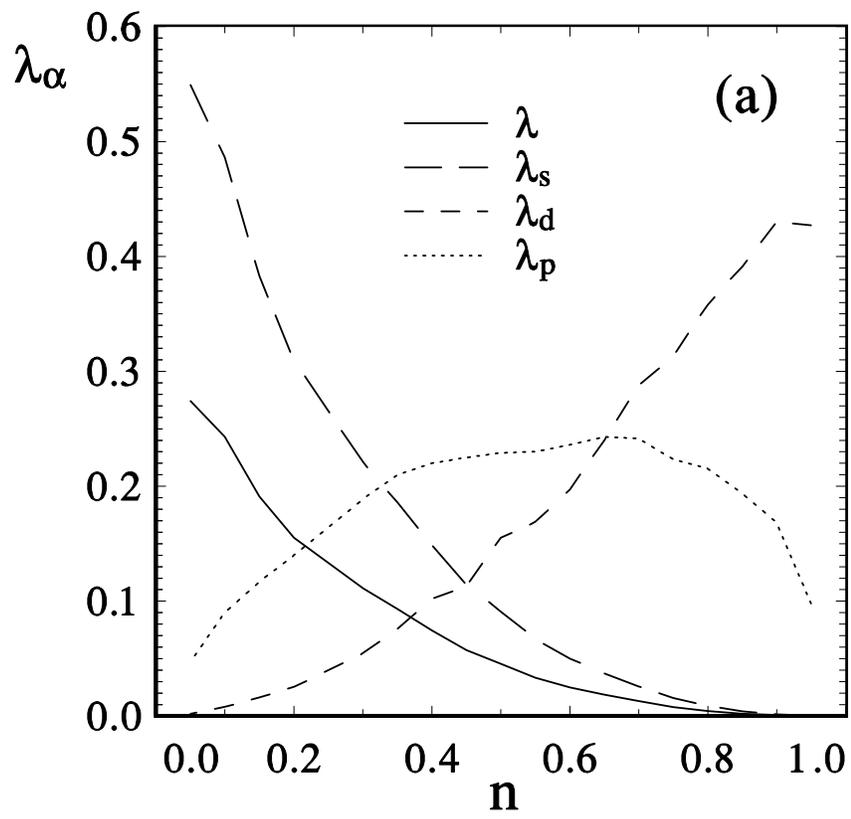
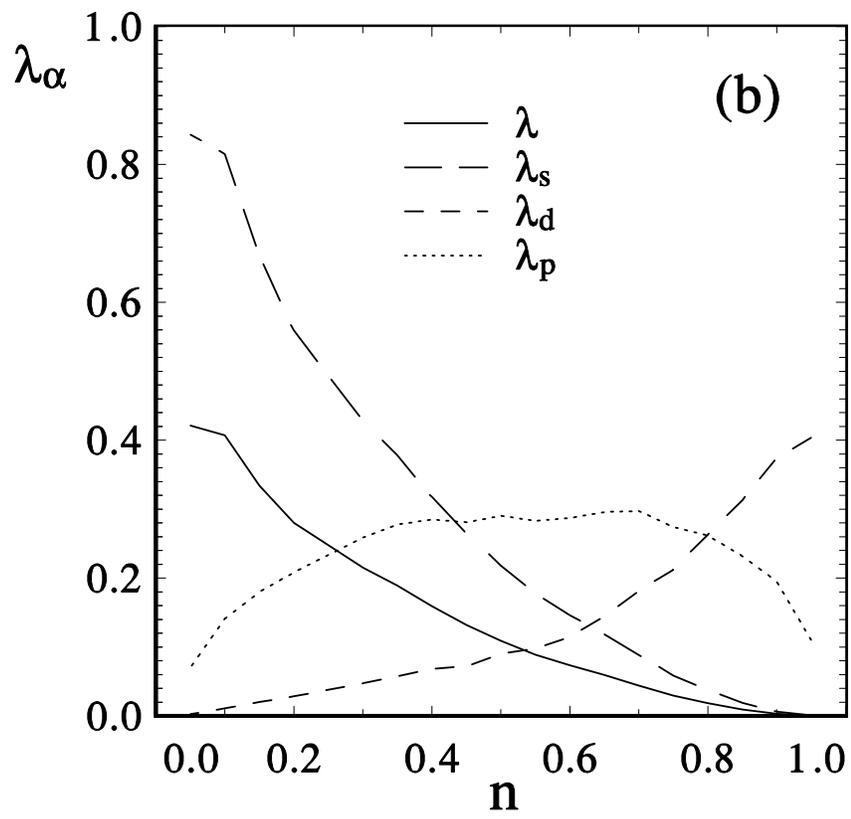
\end{document}